# A technique to avoid Blockchain Denial of Service (BDoS) and Selfish Mining Attack

Md. Ahsan Habib
*Dept. of Computer Science and Engineering*
*Khulna University of Engineering & Technology*
Khulna, Bangladesh
Email: mahabib@cse.kuet.ac.bd

Md. Motaleb Hossen Manik
*Dept. of Computer Science and Engineering*
*Khulna University of Engineering & Technology*
Khulna, Bangladesh
Email: mh.manik@cse.kuet.ac.bd

*Abstract*— Blockchain denial of service (BDoS) and selfish mining are the two most crucial attacks on blockchain technology. A classical DoS attack targets the computer network to limit, restrict, or stop accessing the system of authorized users which is ineffective against renowned cryptocurrencies like Bitcoin, Ethereum, etc. Unlike the conventional DoS, the BDoS affects the system's mechanism design to manipulate the incentive structure to discourage honest miners to participate in the mining process. In contrast, in a selfish mining attack, the adversary miner keeps its discovered block private to fork the chain intentionally that aiming to increase the incentive of the adversary miner. This paper proposed a technique to successfully avoid BDoS and selfish mining attacks. The existing infrastructure of blockchain technology doesn't need to be changed a lot to incorporate the proposed solution.

## I. INTRODUCTION

Bitcoin is a peer-to-peer electronic cash system introduced by Satoshi Nakamoto in 2008 [1]. It has grown into a popular cryptocurrency and its current market capitalization is more than $400B [2]. The data structure behind Bitcoin technology is blockchain. As Bitcoin is a public blockchain, anyone can join to the system anytime. Participants in the blockchain are called nodes and nodes who collect cryptocurrency transactions between clients are known as miners. Miners have to solve a complex mathematical cryptopuzzle to create a block to the blockchain successfully [3]. Every node in the network adds the new transactions *i.e.* block to get updated. A miner needs to spend its resources to solve a complicated mathematical cryptopuzzle and generate Proofs of Work (PoW) [4]. This work is known as mining which is very difficult but easily verifiable. A miner who successfully mines a block gets rewarded with cryptocurrency for his effort.

Bitcoin necessitates a majority of the miners to be truthful. This virtual currency can be controlled by a pool of miners if the pool holds the majority of the mining power called 51% attack that goes against the concept of decentralization. For example, such a pool can forbid a selected transaction or more transactions. It is very common that the Bitcoin miners form pools and behave strategically. For a specific pool, all miners use their mining power to solve the cryptopuzzle and if any miner finds it s/he informs the pool manager and all other miners stop mining that block. If the pool successfully creates a block, the rewards are distributed among the miners of the pool proportionally to their contributions. It decreases the variance of their income rate.

Corresponding Author:
Md. Ahsan Habib
Department of Computer Science & Engineering
Khulna University of Engineering & Technology, Bangladesh
Email: mahabib@cse.kuet.ac.bd

Blockchain Denial of Service (BDoS) is an attack on the blockchain domain that manipulate the reward structure to discourage authentic miners to participate in the process of mining. Another attack on this domain known as selfish mining is a situation when a pool of miners keeps its discovered blocks private to fork the chain intentionally. Both attacks need around one-third of the total mining power to take control of the blockchain. This paper analyzes the above attacks and proposes a solution to avoid these attacks in an efficient manner.

## II. PRELIMINARIES

### A. Blockchain

A Blockchain $B$ is a distributed, decentralized, and tamper-proof storage mechanism consisting of a number of $n$ blocks $b_0, b_1, ..., b_n$. Each block $b_i$ can contain $m$ transactions $tr_0, tr_1, ..., tr_m$ (may vary), a hash of the previous block $h_{i-1}$, a timestamp $TS$, a nonce $r_i$, etc. The hash of the previous block is calculated by $h_{i-1} = hash(b_{i-1})$ and $r_i$ denotes a random number to ensure the validity of the block. The block is formally defined as $b_i = ((tr_0, tr_1, ..., tr_m), h_{i-1}, TS, r_i)$. Blockchain $B$ is thought to be valid if each block $b_i$ is valid where the validity of the $b_i$ depends on the validity of each transaction $tr_i$ in that block and the hash of the $b_i$ is to comply with a certain threshold $th$ [5]. A new block $b_{new}$ gets added to the blockchain $B$ only when the majority of nodes ($> \frac{1}{2}$) in the blockchain have agreed to it by validating all transactions $tr_i$. The mechanism by which the nodes agreed to add the $b_{new}$ is known as the consensus mechanism. PoW is the most renowned consensus mechanism in this domain is discussed in the following step. As new blocks are added to the blockchain, its size continues to grow. A simple $B$ is illustrated in the following Fig. 1. Each $b_i$ in the blockchain has two parts: a block header $b_{ih}$ and a block body $b_{ib}$. Secure Hashing Algorithm (SHA)-256 is a prominent algorithm for hashing used in this domain.

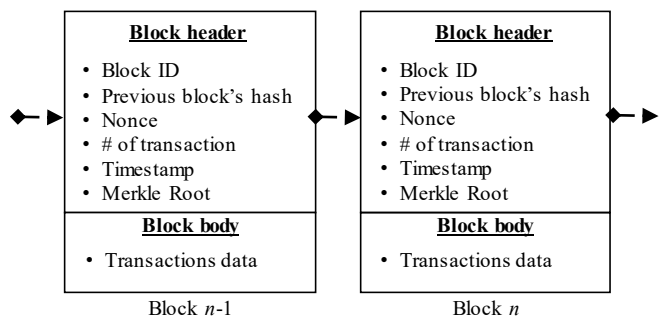

Fig. 1. Blockchain structure.

### B. PoW

Calculating the block header's hash value is the first step in the Proof of Work (PoW) consensus mechanism. The block header $b_{hi}$ contains a $r_i$ that is continually changed by the node participating in the mining process called miner to obtain different hash values of the $b_{hi}$. As the mechanism needs to be found a hash value ($h_i = SHA256(b_i)$) to remain within a certain threshold $th$, miners have to solve a complex mathematical cryptopuzzle. A miner $m_k$ who solves the cryptopuzzle first will add the $b_{new}$ to the $B$. While $m_k$ append the newly created block $b_{new}$ to the $B$, it broadcasts the $b_{new}$ to the blockchain network. The other nodes of the blockchain network validate all the $tr$ of the $b_{new}$ and update the block to its blockchain [4].

### III. DESCRIPTIONS OF ATTACKS

This section demonstrates BDoS and selfish mining attacks in detail.

#### A. BDoS

A Denial of Service (DoS) attack targets the computer network to limit, restrict, or stop accessing the system of authorized users. There have been no successful DoS attacks to date against prominent cryptocurrencies like Bitcoin, Ethereum, etc. Unlike the classical Denial of Service (DoS), Blockchain Denial of Service (BDoS) targets to manipulate the revenue structure to discourage authentic miners to participate in the process of mining.

Let $B^*$ be the current state of the main chain illustrated in Fig. 2(a). An attacker $\check{A}_i$ creates a block $b_{\check{A}}$ and appends it to the $B^*$ and the resultant chain is $B_{\check{A}}$ which is shown in Fig. 2(b). Rather than publishing the entire block $b_{\check{A}}$, the $\check{A}_i$ publishes only the block header $b_{\check{A}h}$ and withholds the block body $b_{\check{A}b}$ that contains mainly the list of transactions $(tr_0, tr_1, ..., tr_t)$. The reason behind publishing only the block header $b_{\check{A}h}$ is to give proof that the $\check{A}_i$ successfully creates the block. At this point, a rational miner $m_k$ can ignore the $b_{\check{A}h}$ of the $\check{A}_i$ and create a block $b_k$ following the main chain $B^*$ and creates the resultant chain $B_k$.

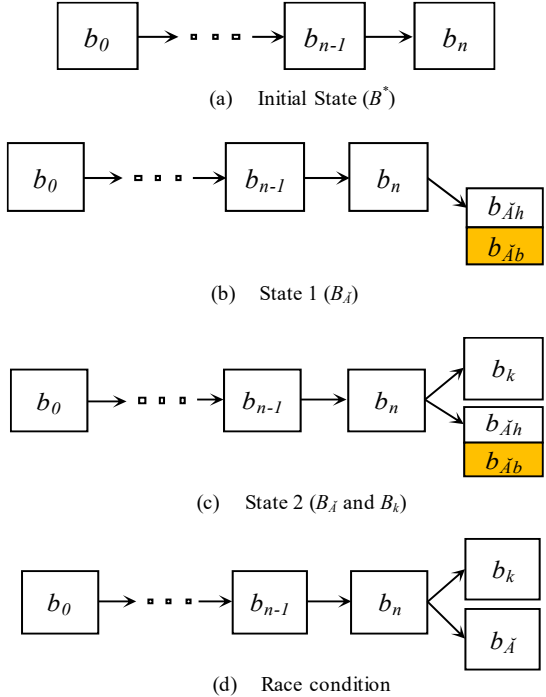

Fig. 2. Blockchain Denial of Service.

As there are two different chains e.g. $B_{\check{A}}$ and $B_k$, this situation is known as a fork that is shown in Fig. 2(c). When the $m_k$ publishes the $b_k$ to the blockchain network, the $\check{A}_i$ also immediately publishes the contents of the $b_{\check{A}}$, resulting in a race between $B_{\check{A}}$ and $B_k$ that is depicted in Fig. 2(d). Depending on the parameters of the system, the block $b_k$ of $m_k$ may or may not get added to the $B^*$. Hence, the probability of getting rewards for the successful block creation of the rational miners $m_{rat}$ lessens. Pausing mining turns out to be a better alternative than mining if the incentive is low enough. If the incentive reduces significantly all the miners stop mining and the resultant situation is that the $\check{A}_i$ can also stop mining after taking the incentive of creating block. The blockchain thus grinds to a complete halt. This attack necessitates significantly lower than 50% power of mining.

#### B. Markov Chain Representation of BDoS

The BDoS problem can be represented in a markov chain model depicted in the Fig. 3. Here, 0, 1, and 2 specify the state number.

***State0:*** specifies initial state of the blockchain (Fig. 2(a)).

***State1:*** denotes the attack on progress and $\check{A}_i$ successfully creates a block and publishes only the $b_{\check{A}h}$ to the blockchain network (Fig. 2(b)).

***State2:*** states that the $m_{rat}$ creates a block ignoring the block of $\check{A}_i$ and starts a race condition (Fig. 2(c) and Fig. 2(d)).

Let the $\check{A}_i$ creates a block $b_{\check{A}}$ in *State0* with its' mining power, $\alpha_{\check{A}}$ and appends it and reaches at *State1*. Rather than publishing the entire block $b_{\check{A}}$, the $\check{A}_i$ publishes only the block header $b_{\check{A}h}$. In Fig. 3(a), it is observed that $m_i$ and $m_{rat}$ are still mining in the *State1* with mining power $\alpha_{m_i}$ and $\alpha_{m_v}$, respectively. Although they know that the probability of rewarding from this state is low. This is true for all the $m_{rat}$. If the $m_i$ stops mining in the *State1* which is depicted in Fig. 3(b), then it will be beneficial for the $\check{A}_i$. Here, $\beta$ ($\leq 1$) represents the rushing ability of the block of $\check{A}_i$ i.e. $\beta.m_{rat}$

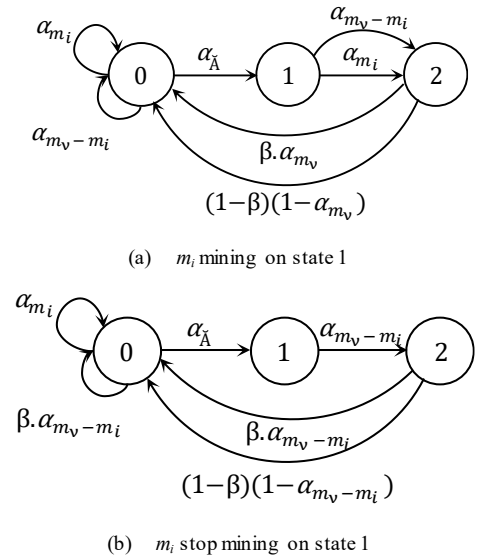

Fig. 3. Markov chains of BDoS.

miners with $\beta.\alpha_{m_v}$ mining power mining on the top of the block of $\breve{A}_i$. In contrast, the rest $(1-\beta).m_{rat}$ miners with $(1-\beta)(1-\alpha_{m_v})$ mining power mining on the top of the block of $m_{rat}$. The reward is given to the winner of the race. This can be shown using the equation (1).

$$R = \begin{cases} \breve{A}_i, & if\ \beta.\alpha_{m_v} > (1-\beta)(1-\alpha_{m_v}) \\ m_{rat}, & if\ (1-\beta)(1-\alpha_{m_v}) > \beta.\alpha_{m_v} \end{cases} \quad (1)$$

In case of Fig. 3(b), the $m_i$ stops mining in *State 1* and it is logical as the block is already created by $\breve{A}_i$. The reward is given to the winner of the race shown in the following equation (2).

$$R = \begin{cases} \breve{A}_i, & if\ \beta.\alpha_{m_v-m_i} > (1-\beta)(1-\alpha_{m_v-m_i}) \\ m_{rat}, & if\ (1-\beta)(1-\alpha_{m_v-m_i}) > \beta.\alpha_{m_v-m_i} \end{cases} \quad (2)$$

*C. Selfish Mining*

Selfish mining is a situation when a pool of selfish miners $m_{self}$ keeps its discovered blocks $b_{self}$ private to fork the chain intentionally. To do it, the selfish mining pool $Pool_{self}$ publishes only selective mined block to invalidate the block $b_{auth}$ of the authentic miners $m_{auth}$. The $Pool_{self}$ retains $b_{self}$ private, secretly forking the blockchain that creates a private branch $br_{pri}$. For the time being, the $m_{auth}$ continue mining on the smaller public branch $br_{pub}$. As the $m_{self}$ hold a comparatively small fraction of the total mining power, the $br_{pri}$ will not lead of the $br_{pub}$ indefinitely. Thus, the $m_{self}$ selectively divulges blocks $b_{self}$ from the $br_{pri}$ to the public, such that the $m_{auth}$ will switch to the recently published blocks, leaving the shorter $br_{pub}$. This makes the block creation labor spent on the shorter $br_{pub}$ wasted, and enables the selfish pool $Pool_{self}$ to collect the greater revenues.

The selfish mining is shown in the following Fig 4. The current state of the main chain is illustrated in Fig 4(a). The strategy is determined by the $Pool_{self}$ or by the other $m_{auth}$. The results depend on the relative lengths the $br_{pri}$ and the $br_{pub}$. They are as follows:

(a) When the $br_{pub}$ is longer than the $br_{pri}$ then the $Pool_{self}$ is behind the $br_{pub}$. As the $m_{self}$ holds less mining power compared to the $m_{auth}$, the chances of the $m_{self}$ mining on the $br_{pri}$ and lead the main branch are low. Therefore, the $Pool_{self}$ must update with the main branch each time its $br_{pri}$ falls behind which is shown in Fig. 4(b).

(b) When the $Pool_{self}$ is able to create a $b_{self}$, it leads a block over the $br_{pub}$ on which the $m_{auth}$ perform mining. At this point, $m_{self}$ keep the $b_{self}$ secret and update its $br_{pri}$ to the $Pool_{self}$ without publishing into the network that is depicted in Fig. 4(c). There may arise two possible outcomes:

  i. Firstly, the $m_{auth}$ discover a $b_{auth}$ on the $br_{pub}$ that cancel out the $Pool_{self}$'s lead. In this case, while the $m_{auth}$ publishes the $b_{auth}$ to the network, the $Pool_{self}$ immediately publishes its $br_{pri}$ containing $b_{self}$, resulting in a race between $br_{pub}$ and $br_{pri}$ which is illustrated in Fig. 4(d) similar to BDoS (Fig. 2(d)). At this point, if the $m_{self}$ manage to create a $b_{self}$, it immediately publishes the $b_{self}$ to win the reward of two blocks which is shown in Fig. 4(e). If the $m_{auth}$ creates a $b_{auth}$ over the mined block of $Pool_{self}$, then the pool wins the reward of its block and the $m_{auth}$ get the profits of its block that is depicted in Fig. 4(f).

  ii. The $Pool_{self}$ mines another $b_{self2}$ and extends its lead two block on the $br_{pub}$ illustrated in Fig. 4(g). At this stage, if the $m_{auth}$ creates a $b_{auth}$, the pool publishes one block from its $br_{pri}$. This process will be continued if the pool has two block or more lead. As the mining power of $m_{self}$ is relatively small compared to other, eventually it turns out to a single block lead. At this point, the pool publishes the $br_{pri}$. As the $br_{pri}$ is larger than $br_{pub}$, all miner adopts it as a main branch. Hence, the pool enjoys all blocks' reward.

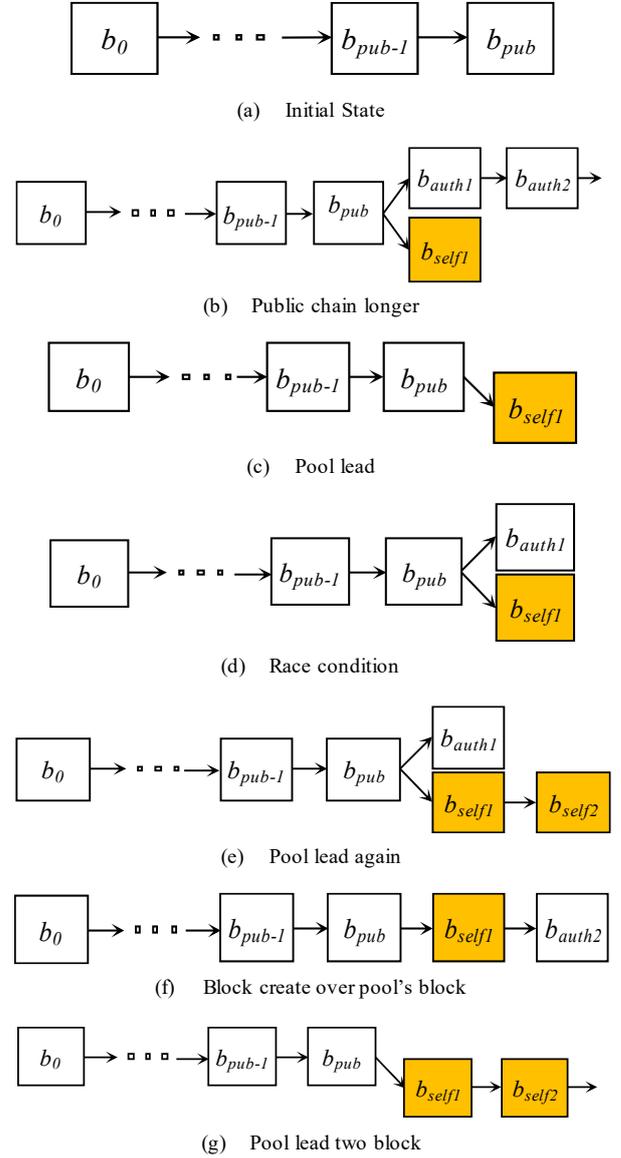

Fig. 4. Selfish Mining.

*D. Markov Chain Representation of Selfish Mining*

The selfish mining problem can be exemplified in a markov chain model depicted in the Fig. 5. Here, 0, 1, 2 and 3 specify the state number of selfish mining.

***State0:*** specifies initial state of the blockchain (Fig. 4(a)).

***State1:*** states that $Pool_{self}$ successfully creates a block $b_{self}$ and keeps it secret (Fig. 4(c)). In this case, pool is on lead.

***State2:*** denotes that the $m_{auth}$ creates a block $b_{auth}$ and cancel out the $Pool_{self}$'s lead. Now the $Pool_{self}$ immediately publishes $b_{self}$, resulting in a race (Fig. 4(d)).

***State3:*** states that the $Pool_{self}$ leads on two block (Fig. 4(g)).

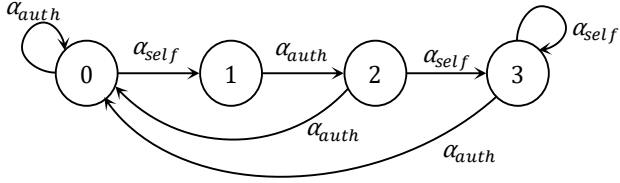

Fig. 5. Markov chain of Selfish Mining.

Let $Pool_{self}$ is in *State0* and creates a blocks $b_{self}$ and now it is in *State1*. It keeps $b_{self}$ confidential to fork the chain intentionally. If $m_{auth}$ discover a $b_{auth}$ that cancel out the $Pool_{self}$'s lead resulting in a race in *State2* as $Pool_{self}$ immediately publishes $b_{self}$ (Fig. 4(d)). The reward is given to the winner of the race. This can be shown using the equation (3).

$$R = \begin{cases} Pool_{self}, & if\ \alpha_{self} > \alpha_{auth} \\ m_{auth}, & if\ \alpha_{auth} > \alpha_{self} \end{cases} \quad (3)$$

In this circumstance, if the $Pool_{self}$ manage to create a $b_{self}$, it immediately publishes the $b_{self}$ to win the reward of two blocks in *State3* (Fig. 4(e)). Otherwise, in *State2* if the $m_{auth}$ discover a $b_{auth}$ over the block of $Pool_{self}$, then both the $Pool_{self}$ and $m_{auth}$ win the reward of their respective block (Fig. 4(f)). As it is a stable situation, the blockchain is in its initial stable state, *State0*. In contrast, if the $Pool_{self}$ mines another $b_{self2}$ and extends its lead of two block then discovering $b_{auth}$ by $m_{auth}$ will not work because if the $m_{auth}$ creates a $b_{auth}$, the pool publishes one block from $br_{pri}$. This process will be continued if the $Pool_{self}$ has two block or more lead presented in *State3*. As the mining power of $m_{self}$ is relatively small, eventually the blockchain turns out to the initial stable state.

## IV. THE PROPOSED SOLUTION

This section briefly describes the proposed system to avoid BDoS attack and to prevent selfish mining problem.

### A. Overview of the proposed solution

The proposed solution mainly works based on adding a dummy block $b_{dummy}$ at the end of the main chain at interval of block creation time $r$ plus overhead $\bar{e}$. Here, $r$ denotes the block generation time for the miners in a blockchain network and $\bar{e}$ denotes some extra time for block propagation, etc. For example, in Bitcoin blockchain the value of $r$ is 10 minutes on overage. Block creation will only take into account the entire block not block header. The proposed technique will automatically create a dummy block $b_{dummy}$ on the existing public main branch, $br_{pub}$ to avoid from the BDoS attack and selfish mining attack. Hence, the blockchain will be more stabilized and all the miners/pools will get equal chance for creating blocks. But in some cases, there may have some loss of the honest miners because of time limit but this is negligible. Although the proposed solution allows BDoS and selfish mining within $(r + \bar{e})$ time but the probability of happening this is very low.

### B. Avoidance of BDoS

Let, the attacker $\breve{A}_i$ creates a block $b_{\breve{A}}$ and appends it to the $B^*$ and the resultant chain is $B_{\breve{A}}$. Rather than publishing the entire block $b_{\breve{A}}$, the $\breve{A}_i$ publishes only the block header $b_{\breve{A}h}$ and withholds the block body $b_{\breve{A}b}$ that contains mainly the list of transactions $(tr_0, tr_1, ..., tr_t)$. At some point, the system has been passed the $(r+\bar{e})$ times, the system automatically generates an dummy block on the public main chain, $br_{pub}$. As $(r+\bar{e})$ times have been passed, the block header $b_{\breve{A}h}$ and withholds the block body $b_{\breve{A}b}$ are discarded and the $b_{dummy}$ is added. The entire process is depicted in the following Fig. 6.

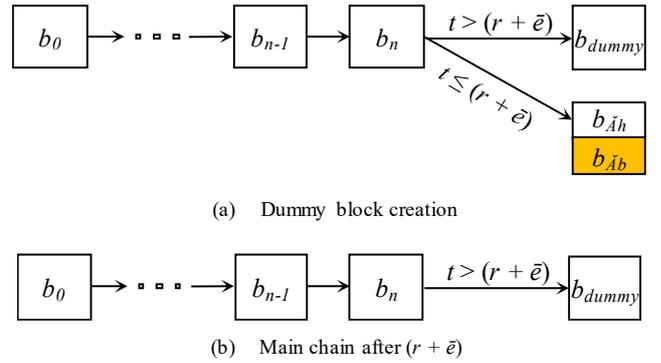

(a) Dummy block creation

(b) Main chain after $(r + \bar{e})$

Fig. 6. Proposed solution to solve BDoS.

### C. Avoidance of selfish mining

Let, the pool $Pool_{self}$ creates a new block $b_{self}$, it leads a block over the $br_{pub}$ on which the $m_{auth}$ perform mining. At this point, if $m_{self}$ keeps the $b_{self}$ secret and update its $br_{pri}$ to the $Pool_{self}$ without publishing into the network, it is allowed only within the $(r+\bar{e})$. After ending the period $(r+\bar{e})$, the system automatically generates a dummy block on the public main chain *i.e.* the $b_{dummy}$ is added to $br_{pub}$. No new block $b_{new}$ is accepted on the previous chain. Fig. 7. depicts the entire process.

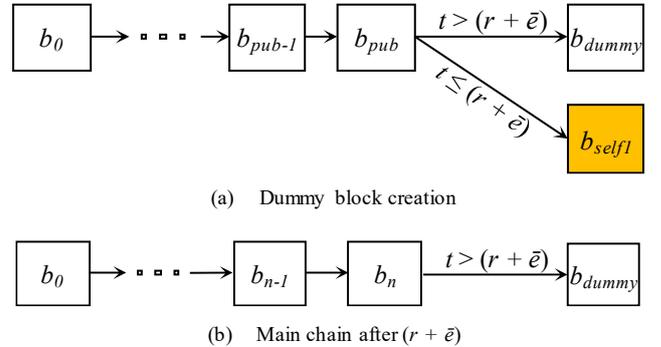

(a) Dummy block creation

(b) Main chain after $(r + \bar{e})$

Fig. 7. Proposed solution to solve Selfish Mining.

## V. SOLUTION ANALYSIS

The proposed solution described above successfully avoids the BDoS attack and prevents the selfish mining problem. However, the solution has some limitations also.

The analysis of the proposed solution described in the following.

### A. Loss of honest miner

Let, honest miners are trying to create a block with some transactions but within $(r + \bar{e})$ time they don't manage to append the block to the blockchain. As dummy block $b_{dummy}$ is added to the main chain after $(r + \bar{e})$ time, the effort of the honest miners will be lost fully. Let $\varrho$ is the probability that the honest miners are not able to mine of a block then the amount of loss of honest miners in BDoS attack can be expressed by the following equation (4).

$$Loss_{miners} = \varrho \times \alpha_{m_v} \quad (4)$$

### B. Loss of attacker

Let, the attacker is not able to manage to append the block to the blockchain with probability $z$ within $(r + \bar{e})$ time. As dummy block $b_{dummy}$ is added to the main chain, the effort of the attacker will be lost fully. The amount of loss of attacker in BDoS attack can be expressed by the equation (5).

$$Loss_{attacker} = z \times \alpha_{\check{A}} \quad (5)$$

### C. BDoS Attack within $(r + \bar{e})$ time

The proposed solution allows BDoS within $(r + \bar{e})$ time but in such scenario is very rare. Let, the attacker $\check{A}_i$ creates a block within $(r + \bar{e})$ time and only publishes the block header to discourage honest miners to participate in the mining process. The honest miners still mining on the top of the main chain knowing that the dummy block may be added after $(r + \bar{e})$ time discarding the block of attacker if it is not fully published. But if the attacker is able to create another block within the $(r + \bar{e})$ time and publishes the first block fully then the efforts of honest miners will be totally lost. Now the attacker has another $(r + \bar{e})$ time for creating another block. If it is able to find another block, then it may publish the previous block and gain the rewards. This process may be continuing until the attacker is able to create a block within $(r + \bar{e})$ time. But for creating two blocks within $(r + \bar{e})$ time needs high computation cost which is inefficient to the attacker. In contrast, dummy block adding time $(r + \bar{e})$ needs to adaptive to mitigate this scenario. Thus probability of BDoS attack against the proposed solution is very low.

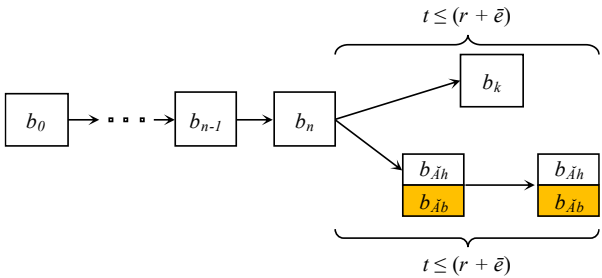

Fig. 8. BDoS Attack within $(r + \bar{e})$ time.

### D. Selfish mining within $(r + \bar{e})$ time

The proposed solution allows selfish mining within $(r + \bar{e})$ time but in such scenario is very rare. Let, the pool $Pool_{self}$ creates a block within $(r + \bar{e})$ time keeps it secret. Again if the pool $Pool_{self}$ is able to create another block within the $(r + \bar{e})$ time and publishes the first block only then the efforts of honest miners will be totally lost. Now the attacker has $(r + \bar{e})$ time for creating another block. If it is able to find another block, then it may publish the previous block and gain the rewards. This process may be continuing until the attacker is able to create a block within $(r + \bar{e})$ time. But for creating two blocks within $(r + \bar{e})$ time needs high computation cost which is inefficient to the attacker. In contrast, dummy block adding time $(r + \bar{e})$ needs to adaptive to mitigate this scenario. Thus probability of selfish mining against the proposed solution is very low.

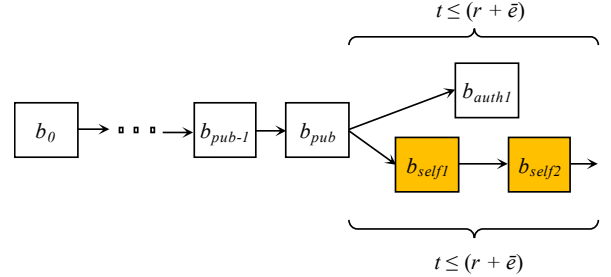

Fig. 9. Selfish mining within $(r + \bar{e})$ time.

### E. Race between dummy block and new block

For the usual cases, the dummy block propagation time over the blockchain network is assumed to be $\bar{e}$. For some cases, it may be more than the $\bar{e}$ due to different constraints. Then there is a possibility of race condition if some of miners manage to create a block before getting the dummy block. This scenario is depicted in the following Fig. 10(a). However, the problem of forking in this case is trouble-free as the $(r + \bar{e})$ time is over, no new block is allowed to the main chain except the dummy block. So, the main chain after $(r + \bar{e})$ time is depicted in the Fig. 10(b).

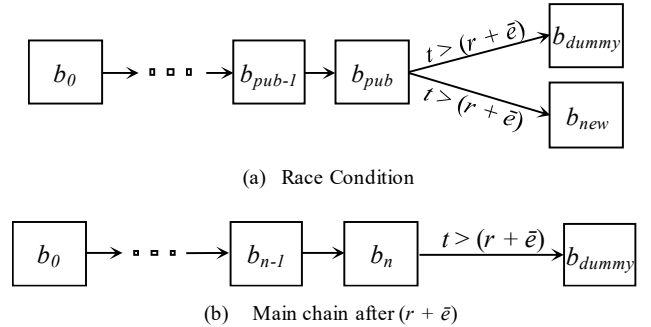

(a) Race Condition

(b) Main chain after $(r + \bar{e})$

Fig. 10. Forking after $(r + \bar{e})$ time.

## VI. CONCLUSION

Blockchain denial of service (BDoS) and selfish mining are the two most critical attacks on blockchain technology which may affect incentive structure and increase the incentive of adversary miner, respectively. Unlike the conventional DoS, the BDoS affects the system's mechanism design to manipulate the incentive structure to discourage honest miners to participate in the mining process. In contrast, in a selfish mining attack, the adversary miner keeps its discovered block private to fork the chain intentionally that aiming to increase the incentive of the adversary miner. This paper proposed a technique to successfully avoid BDoS and selfish mining attacks. The existing infrastructure of blockchain technology doesn't need to be changed a lot to incorporate the proposed solution.